\documentclass[preprint,showpacs]{revtex4}
\usepackage{graphicx}
\usepackage{dcolumn}
\usepackage{amsmath}
\usepackage{bm}
\newcommand{\be}{\begin{equation}}
\newcommand{\ee}{\end{equation}}
\newcommand{\ba}{\begin{eqnarray}}
\newcommand{\ea}{\end{eqnarray}}
\begin{document}
\title{Systematic Improvement of Classical Nucleation Theory}
\author{Santi Prestipino$^1$~\cite{aff1}, Alessandro Laio$^2$~\cite{aff2},
and Erio Tosatti$^{2,3}$~\cite{aff3}}
\affiliation{
$^1$ Universit\`a degli Studi di Messina, Dipartimento di Fisica,
Contrada Papardo, I-98166 Messina, Italy \\
$^2$ International School for Advanced Studies (SISSA) and CRS Democritos,
CNR/INFM, Via Bonomea 265, I-34136 Trieste, Italy \\
$^3$ The Abdus Salam International Centre for Theoretical Physics (ICTP),
P.O. Box 586, I-34151 Trieste, Italy}
\date{\today}
\begin{abstract}
We reconsider the applicability of classical nucleation theory (CNT)
to the calculation of the free energy of solid cluster formation in a
liquid and its use to
the evaluation of interface free energies from nucleation barriers.
Using two different freezing transitions (hard spheres and NaCl)
as test cases, we first observe that the interface-free-energy estimates
based on CNT are generally in error.
As successive refinements of nucleation-barrier theory, we consider
corrections due to a non-sharp solid-liquid interface and to a non-spherical
cluster shape. Extensive calculations for the Ising model show that
corrections due to a non-sharp and thermally fluctuating interface account
for the barrier shape with excellent accuracy.
The experimental solid nucleation rates that are measured in colloids are
better accounted for by these non-CNT terms, whose effect appears to be
crucial in the interpretation of data and in the extraction of the
interface tension from them.
\end{abstract}
\maketitle

The decay of metastable states, such as the solidification of a supercooled
liquid, takes place through the nucleation and growth of
some small-sized droplet within the system~\cite{Kelton}. 
The initial stage of the phase transformation is usually described within the
time-honored classical nucleation theory (CNT)~\cite{Volmer,Farkas,Becker},
where the droplet is envisaged as a sphere of, say, bulk solid, separated
from the liquid by a sharp interface, giving rise to a free-energy penalty
proportional to the interface area and a total Gibbs-free-energy activation
barrier
\be
\Delta G(n)=-|\Delta\mu|n+An^{2/3}\,,
\label{eq1}
\ee
where $n$ is the number of particles in the solid cluster, $\Delta\mu<0$
is the chemical potential difference between solid and liquid,
$A=(36\pi)^{1/3}\rho_s^{-2/3}\sigma$ with $\rho_s$ the bulk-solid number
density and $\sigma$ the specific surface energy (surface tension) of the
planar interface, all anisotropies being neglected at this stage.
The droplet grows if it exceeds a critical size $n^*$ corresponding to the
maximum $\Delta G(n)$ ($\equiv\Delta G^*$). CNT is routinely used to estimate
the nucleation rate $I=I_0e^{-\beta\Delta G^*}$, where $\beta=1/(k_BT)$ is
the inverse temperature and $I_0$ a kinetic prefactor that varies slowly
with $T$. Clearly, this connection between $I$ and $\sigma$~\cite{Becker}
relies on several severe approximations. First of all the choice of an
appropriate reaction coordinate, here the droplet size $n$, an issue
largely discussed and criticized in the
literature~\cite{Chandler,Dellago,Moroni,Lechner}.
Moreover, $I_0$ is notoriously influenced by genuinely non-equilibrium
effects and various expressions resulting from a more detailed consideration
of the nucleation kinetics are known since a long
time~\cite{Zeldovich,Frenkel,Langer}. 

In this Letter, we do not address the issue of the validity of
CNT for predicting the nucleation rate but rather consider an even more
fundamental question, namely the efficacy of CNT in describing the dependence
of the interface free energy of the solid cluster on its size. Our starting
point is to show that the profiles of $\Delta G(n)$ obtained by numerical
simulation of nucleation clusters in a variety of systems are not consistent
with Eq.\,(\ref{eq1}). We then explore corrections, some already present in
the literature, some novel. It emerges that the numerical profiles can be
accurately reproduced by assuming a diffuse and thermally fluctuating
solid-liquid interface.
Finally, we show how this finding is of direct use to interpret nucleation
rates and correctly extract interface free energies from them, a result
that should be of considerable interest to experimentalists.

%
%
\begin{figure}
\includegraphics[width=7cm]{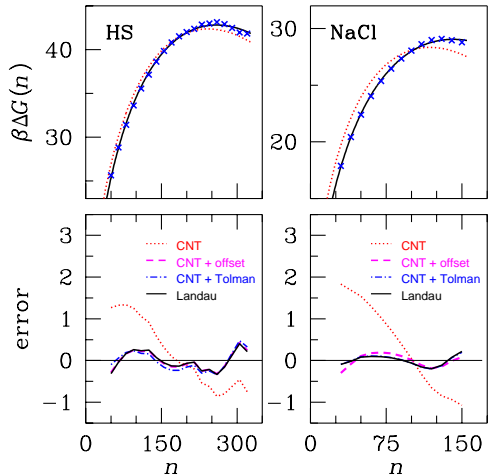}
\caption{(Color online).
Gibbs-free-energy cost $\Delta G(n)$ of an $n$-cluster in units of $k_BT$.
Left: hard spheres at a packing fraction of 0.5207, from~\cite{Auer};
right: NaCl at $T=825$\,K, from~\cite{Valeriani}.
Top panels: blue crosses, MC data (selection of data points);
red dotted line, CNT best fit; black solid line, Landau-theory best fit.
In all fits, data for $n\lesssim n^*/5$ are ignored.
Bottom panels: deviation of the fitting curves from the data.
CNT ($\widetilde{\delta}=\widetilde{\epsilon}=C=0$ in Eq.\,(\ref{eq2}));
CNT + offset ($\widetilde{\delta}=C=0$);
CNT + Tolman ($\widetilde{\epsilon}=C=0$); Landau ($C=0$).
Values for $\sigma$ from CNT are: hard spheres, $\beta\sigma d^2=0.724$
(sphere diameter $d$); NaCl (cubic nucleus), $\sigma=79.75$
erg/cm$^2$. In hard spheres, the optimal $\beta\widetilde{\sigma}d^2$ is
0.737 for Landau theory (with $\widetilde{\delta}=-0.017\,d$ and
$\widetilde{\epsilon}=-0.332\,d^2$), 0.741 for CNT + offset (with
$4\pi\beta\widetilde{\sigma}\widetilde{\epsilon}=-2.599$),
and 0.761 for CNT + Tolman (with $\widetilde{\delta}=0.086\,d$).
In NaCl (cubic nucleus), the optimal $\widetilde{\sigma}$ is 88.87
erg/cm$^2$ for Landau theory (with $\widetilde{\delta}=0.716$\,\AA\, and
$\widetilde{\epsilon}=0.471$\,\AA$^2$), 83.74 erg/cm$^2$ for CNT + offset
(with $6\beta\widetilde{\sigma}\widetilde{\epsilon}=-3.833$), and 88.53
erg/cm$^2$ for CNT + Tolman (with $\widetilde{\delta}=0.675$\,\AA).
Note the large improvement over CNT obtained with just one more fitting
parameter and how similar is the quality of the fit based on CNT + offset
to the Landau fit.}
\label{fig1}
\end{figure}

We begin by displaying in Fig.\,1 existing accurate simulation data for
$\Delta G(n)$ of a solid cluster nucleating inside a bulk liquid,
available for hard spheres~\cite{Auer} and for the Fumi-Tosi model of
NaCl~\cite{Valeriani}. In each case, we superpose a CNT least-square fit to
Eq.\,(\ref{eq1}) for comparison. It is clear that CNT is not generally
adequate to describe $\Delta G(n)$. The deviations are systematic and of
different sign at low and large $n$. The fit quality does
not improve by restricting data to large clusters only, indicating that even
in the barrier region the cluster free energy does not obey Eq.\,(\ref{eq1}).
To shed light on this failure of CNT, we relax the approximations leading
to Eq.\,(\ref{eq1}) one at a time. Eventually, we shall get a more general
expression for the free-energy cost of a $n$-particle cluster for large $n$,
which turns out to have the Dillmann-Meier~\cite{Dillmann} form
\be
\Delta G=4\pi R^2\widetilde{\sigma}\left(1-\frac{2\widetilde{\delta}}{R}+
\frac{\widetilde{\epsilon}}{R^2}\right)-\frac{4}{3}\pi R^3\rho_s|\Delta\mu|+
C\ln\frac{R}{a}\,,
\label{eq2}
\ee
with $R=[3n/(4\pi\rho_s)]^{1/3}$ and $a$ a microscopic length, and where
$\widetilde{\sigma},\widetilde{\delta},\widetilde{\epsilon}$, and $C$
are theory-dependent parameters.
The assumption in (\ref{eq2}) is that of a {\em spherical} cluster
shape --- a different shape, that would be determined by free-energy anisotropy,
would slightly change the value of $\widetilde{\sigma}$ but not the physical
discussion that follows. The first modification to CNT we consider is
dropping the sharp-interface approximation~\cite{Cahn}. Within Landau
theory, the free-energy cost of the critical droplet is the unstable
stationary point of a functional, e.g.
\be
{\cal G}[\phi]=\int{\rm d}^3x\,\left\{\frac{c}{2}(\nabla\phi)^2+
\frac{\kappa}{2}(\nabla^2\phi)^2+g(\phi({\bf x}))\right\}\,.
\label{eq3}
\ee
Here $c,\kappa>0$, $\phi({\bf x})$ is the ``crystallinity'' order parameter
(OP) that distinguishes the solid ($\phi>0$) from the liquid ($\phi=0$), and
$g(\phi)$ is the Landau free energy per unit volume of the homogeneous system.
Below melting, $T<T_m$, $g$ shows, besides the liquid minimum
$g(0)=0$, a second and deeper solid minimum. Right at $T_m$, we assume
$g(\phi)=c_{20}\phi^2(1-\phi/\phi_{s0})^2(1+\tau\phi/\phi_{s0})$
with $c_{20}>0$ and $\tau>-1$~\cite{nota}, where $\phi_{s0}$ is the value
of $\phi$ in the bulk solid at coexistence, and where a non-zero value of
$\tau$ creates an asymmetry between the liquid and the solid minimum.
We also assume that, slightly below $T_m$ and at fixed pressure, $g$ acquires
a linear dependence on $\Delta T=T-T_m$ only through its $\phi^2$ term,
which becomes $c_2\phi^2$ with $c_2=c_{20}+c'_{20}\Delta T$.
With this standard setup, the free energy of a cluster of radius $R$ is
${\cal G}[\phi_R]$, where $\phi_R(r)$ is the spherically-symmetric OP profile
of the cluster. Assuming, as in \cite{Fisher}, that for small supersaturation
and large $R$, $\phi_R(r)$ may be approximated with $\phi_0(r-R)$, where
$\phi_0(z)$ is the OP profile for a planar interface centered at $z=0$, the
cluster free energy takes precisely the form (\ref{eq2}), with $C=0$, 
$\rho_s\Delta\mu=c'_{20}\phi_{s0}^2\Delta T$, and
$\widetilde{\sigma},\widetilde{\delta},\widetilde{\epsilon}$ all linear
functions of $\Delta T$, expressed in terms of $c,\kappa$, and
$\phi_0(z)$~\cite{noi}.
At coexistence and to first order in the deviations from the $\phi^4$ theory
(viz. $\kappa=\tau=0$), we have for $\widetilde{\sigma}(T_m)\equiv\sigma$, etc.
\ba
\sigma&=&\frac{c\phi_{s0}^2}{3\ell}\left(1+\frac{1}{4}\tau+
\frac{2}{5}\,\frac{\kappa}{c\ell^2}\right)\,,\,\,
\delta=\frac{5\ell}{48}\tau\,,\,\,\,{\rm and}
\nonumber \\
\epsilon&=&\ell^2\left[\frac{\pi^2-6}{12}\left(1-\frac{\tau}{4}\right)+
\left(\frac{26}{5}-\frac{\pi^2}{3}\right)\frac{\kappa}{c\ell^2}\right]\,,
\label{eq4}
\ea
where $\ell=\sqrt{2c/c_{20}}$ is a measure of the interface width.

A second effect that is absent in CNT but present in nature and observed
in simulations is that shapes of clusters, far from being static, fluctuate
widely away from their mean shape~\cite{Filion,Zykova}. 
To describe shape fluctuations, we employ a 
field theory for the Canham-Helfrich (CH) Hamiltonian, containing
spontaneous-curvature and bending-energy terms in addition to surface
tension. A CH interface Hamiltonian ${\cal H}_s$ can be derived from the
free-energy functional (\ref{eq3}) for small deviations of the interface
from planarity. Denoting by $\Sigma$ the generic closed-surface profile
and by $\hat{\bf n}$ its outward normal, we obtain~\cite{noi}
\be
{\cal H}_s=\int_\Sigma{\rm d}S\left(\sigma-\sigma\delta\,
\nabla\cdot\hat{\bf n}+\frac{1}{2}\lambda
\left(\nabla\cdot\hat{\bf n}\right)^2\right)\,,
\label{eq5}
\ee
where $\sigma$ and $\delta$ are the same as in Landau theory and
$\lambda=\kappa\phi_{s0}^2/(3\ell)$ under the same hypotheses for which
Eq.\,(\ref{eq4}) holds.
$\Delta G(R)$ can be evaluated explicitly~\cite{noi} for a {\em quasispherical}
cluster~\cite{Milner}, where only quadratic deviations from sphericity are
kept. The wavelength of surface undulations is cut off at a lower limit
$a=\rho_s^{-1/3}$ to account for the granularity of matter.
We find that the surface free energy has a form consistent with
Eq.\,(\ref{eq2}), with new $T$-dependent parameters
$\widetilde{\sigma},\widetilde{\delta}$, and $\widetilde{\epsilon}$
(whose explicit expressions are given in \cite{noi}) and with $C=-(7/3)k_BT$,
which shows that small deviations around a
nominally spherical cluster shape simply add a universal logarithmic correction
to the mean-field functional form of $\Delta G$. This correction is responsible
for the well known $R^{*7/3}$ term in the exponential prefactor of the
nucleation rate~\cite{Guenther}.

Clearly, the parameters in Eq.\,(\ref{eq2}) are determined by
the values of $c$ and $\kappa$ in (\ref{eq3}), as well as by the form of
$g(\phi)$ --- all system-dependent quantities that require a case-specific
theory. We here aim at elucidating the relative importance of the different
terms $\widetilde{\delta},\widetilde{\epsilon},C$ implied by interface
thickness and shape fluctuations. To get a quantitative measure of that,
we directly fit the parameter values in (\ref{eq2}) to the numerical results
for $\Delta G(n)$ for the two systems of Fig.\,1. Consistently with the
assumptions underlying our mesoscopic description, each fit is made only
to data points for sufficiently large $n$. We first include the leading
$\propto\widetilde{\delta}n^{1/3}$ (``Tolman''~\cite{Tolman}) correction to
CNT. As shown in Fig.\,1, this improves the quality of the fit significantly.
The error is reduced substantially in both systems,
although not monotonically. Only a marginal improvement is obtained if both
$\widetilde{\delta}$ and $\widetilde{\epsilon}$ are allowed in the fit.
The inclusion of the logarithmic shape correction gives no
further appreciable gain. Next, we attempted fitting the data by retaining
just the offset ($\equiv 4\pi\widetilde{\sigma}\widetilde{\epsilon}$) in
(\ref{eq2}) beyond $\widetilde{\sigma}$. Alone, the simple offset gave an
improvement of about the same quality as with all terms allowed. We conclude
that corrections to CNT exclusively deriving from a fluctuating cluster shape
appear to be much smaller than those arising, already in Landau theory, from
allowing a non-zero thickness of the interface (shape fluctuations are not
anyway immaterial since they renormalize, even significantly, the
Landau-theory parameters~\cite{noi}). Moreover, either the
Tolman correction or, alternatively, the constant offset each lead to
significant fit improvement over CNT.
The origin of both terms is in the finite thickness of the interface, which
makes the reversible work to create a cluster systematically smaller than
what would be needed for the same cluster with a sharp spherical interface.

%
%
\begin{figure}
\includegraphics[width=7cm]{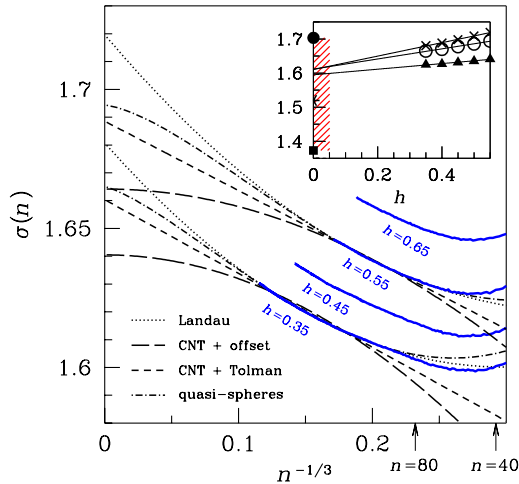}
\caption{(Color online).
The cluster interface free energy of the 3D Ising model in units of $J/a^2$
($a$ being the cubic-lattice spacing), plotted as a function of $n^{-1/3}$
for various $h$ values. The temperature is $T=0.6\,T_c$; starting at $T$
with all spins down, the system is quenched to $h$.
Two spins are part of the same cluster if there is an uninterrupted chain
of up spins between them.
The lattice consisted of $20^3$ sites; for $h=0.35$ a calculation on
$25^3$ sites led to practically the same $\Delta G(n)$.
Three different instances of umbrella potential were considered and all led
to the same $\Delta G(n)$ to within less than $0.1\,k_BT$~\cite{footnote}.
Thick blue lines, MC data for $\Delta G(n)$; black lines, least-square
fits of the $n>80$ data points for $h=0.35$ and $0.55$, based on various
extensions of the CNT (see legend). Note that only the full Landau
expansion captures the upward curvature of $\Delta G(n)$, and that 
especially the shape fluctuations capture
that of smaller clusters. In the inset, different
ways of extracting the interface tension $\sigma$ out of finite-$h$ values
of $\sigma(\infty)$ are compared: triangles, CNT; crosses, Landau theory
($C=0$ in Eq.\,(\ref{eq2})); open dots, quasispheres ($C=-(7/3)k_BT$).
Linear extrapolation of data points at $h=0$ yields
$\sigma\approx 1.60$. The black dot is the value of $\sigma$ calculated 
for the (001) interface (from Ref.\,\cite{Hasenbusch}). The red shading
indicates $\sigma$ estimates from Eq.\,(\ref{eq6}) for cluster shapes
intermediate between cubic (lower end) and spherical (upper end).
Gratifyingly, the $\widetilde{\sigma}$ values extrapolate as they should
to a $\sigma_0$ which is higher than that of the (001) interface, and
intermediate between cubic and spherical shape - the average shape being
also intermediate between the two.} 
\label{fig2}
\end{figure}

The existing simulation data do not permit to
assess the relative importance of the two smooth-interface contributions
$\widetilde{\delta}$ and $\widetilde{\epsilon}$ and of the logarithmic
correction; more specific work is
needed in order to decide that case by case. Using the 3D Ising model
as a test system, we carried out extensive simulations
at moderate supersaturations, computing the cluster free energy
for the nucleation process of magnetization reversal
by the same method as in Refs.~\cite{tenWolde,Bowles,Pan,Maibaum}. We computed
$\Delta G(n)$ for a number of values of the field $h$
($0.35,0.40,\ldots,0.65$, in $J$ units) and plotted the ratio $\sigma(n)$
of the surface free energy $F_s=\Delta G(n)+|\Delta\mu|n$ to the area
$(36\pi)^{1/3}(na^3)^{2/3}$ of the cluster surface as a function of the
inverse radius $n^{-1/3}$ (see Fig.\,2). We verified that, for all
$h$ values considered, clusters close to critical indeed contain the vast
majority of up spins in the system, coherently with the physical picture
at the basis of our theories. It is evident that only the
joint consideration of $\widetilde{\delta}$ and $\widetilde{\epsilon}$
is able to reproduce the upward concavity of $\sigma(n)$ as a function
of $n^{-1/3}$ in the $n$ region ($n>40$) where $F_s(n)\propto n^{2/3}$.
A positive offset $\widetilde{\epsilon}>0$ is confirmed, as expected from
Landau-theory results for $\epsilon$ and from the formula for $\lambda$ in
Eq.\,(\ref{eq5})~\cite{noi}.
In our regime of $h$, the logarithmic term does not change the quality
of the fit; as shown in Fig.\,2, this correction becomes sizeable only
at values of $n$ outside the fit range. However, inclusion
of the logarithm has consequences on the optimal $\widetilde{\delta}$ values,
which reduce from $\approx 0.10$ to $\approx 0.02$ throughout the $h$ range
considered (the Tolman length $\delta$~\cite{Tolman} is zero for
the Ising model at coexistence~\cite{Fisher}).
From this example we conclude that a) neither $\widetilde{\delta}$ nor
$\widetilde{\epsilon}$, both arising from the finite interface width, can 
generally be neglected in the description of the nucleation free-energy
barrier; b) shape fluctuations improve the description especially for
small cluster sizes ($n<80$).

Far from being academic, the existence of these corrections to
CNT has a direct impact on the understanding of experiments,
in particular on the all-important extraction of the interface free energy
$\sigma$ from measured nucleation rates. Assuming the standard activated
expression for $I$, $\sigma$ can be extracted from the slope of
$Y=\ln(I/I_0)$ as a function of $X=(T_m/\Delta T)^2$~\cite{Turnbull}.
If CNT were exact, this slope would be a constant throughout the region of
liquid metastability. When the more general Eq.\,(\ref{eq2}) is employed for
$\Delta G^*$, the slope depends on the distance from coexistence $\Delta T$,
as demanded by non-zero values of
$\widetilde{\sigma},\widetilde{\delta},\widetilde{\epsilon},C$ and their
rates of variation with $\Delta T$. Close to coexistence, one can write
$Y(X)$ as a power series in $\Delta T$:
\be
\ln(I/I_0)=-\frac{\alpha T_m^2}{\Delta T^2}-\frac{\alpha'T_m}{|\Delta T|}+
{\cal O}(1)\,.
\label{eq6}
\ee
Here $\alpha$ takes the same value as in CNT, 
$\alpha=16\pi\sigma^3/(3k_BT_m\rho_s^2L_m^2)$ with $L_m$ the latent heat of
melting per particle. However, $\alpha'$ is not universal:
\be
\alpha'=\alpha\left(1+\frac{3\sigma'T_m}{\sigma}-
\frac{3\rho_sL_m\delta}{\sigma}\right)\,,
\label{eq7}
\ee
taking $\widetilde{\sigma}=\sigma+\sigma'|\Delta T|+\ldots$ close to
coexistence. For instance, for the $\phi^4$ theory it turns out that
$\alpha'/\alpha=1+3\rho_s\ell L_m/(2\sigma)>0$.

%
%
\begin{figure}
\includegraphics[width=7cm]{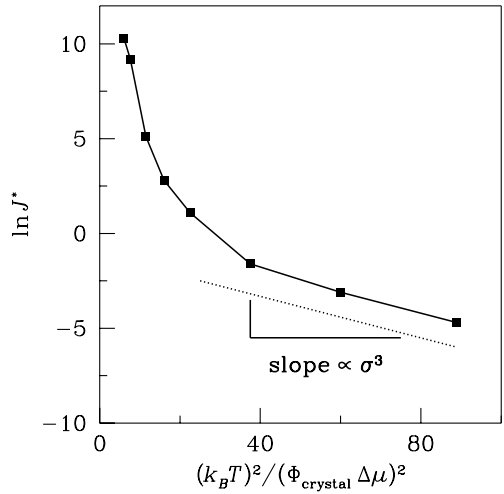}
\caption{Nucleation-rate data for the solidification of a colloidal fluid
(Fig.\,7b of \cite{Franke}), showing $Y=\ln(I/I_0)$ as
a function of a quantity akin to $X=(T_m/\Delta T)^2$ ($\Phi_{\rm crystal}$
is the solid volume fraction). The dotted line gives the slope from which
the surface tension should be extracted.}
\label{fig3}
\end{figure}

Due to (\ref{eq6}), $Y(X)$ develops a concavity, which is upward if
$\alpha'>0$, as is the case for example in colloids (see below).
The very important practical consequence is that the solid-liquid interface
free energy at coexistence ($\sigma$), the key quantity which one wishes
to extract from nucleation rates, {\em is determined by the slope}
($\alpha$) {\em of $Y(X)$ at asymptotically large $X$ and not from the slope,
generally different, at small $X$}. As an example, deviations from linearity
in the $Y(X)$ plot are experimentally evident in colloids, see e.g.
Refs.~\cite{Palberg,Franke}. Fig.\,3 shows how
data should be read to extract $\sigma$.
Since this procedure is not to our knowledge universally followed, this
suggests that at least some tabulated $\sigma$ values may need a revision.
Because the finite-interface corrections reduce the barrier height compared
to CNT, it is to be expected that the true interface free energies are
substantially smaller than believed so far.

We gratefully acknowledge C. Valeriani and S. Auer for sending us
their MC data, and a discussion with G. Parisi. This project was
co-sponsored by CNR through ESF Eurocore Project FANAS AFRI, by the
Italian Ministry of Education and Research through PRIN COFIN Contract
20087NX9Y7, and by SNF Sinergia Project CRSII2\_136287/1.

\newpage
\large
\begin{center}
{\bf Supplementary Material}\\
Santi Prestipino, Alessandro Laio, and Erio Tosatti
\end{center}
\vspace{5mm}
\normalsize

%
%
\section{Landau theory of nucleation}
\setcounter{equation}{0}

In the following, we shall refer to the main text of the manuscript as MT.

A Landau theory of nucleation based on a Cahn-Hilliard-like
functional~\cite{SMCahn}
describing the free-energy cost of a diffuse interface between two phases
--- ``solid'' and ``liquid'', neglecting anisotropies --- can be formulated
as follows.
Assume for simplicity a scalar order-parameter (OP) field $\phi({\bf x})$.
The Landau free energy is (Eq.\,(3) of MT):
\be
{\cal G}[\phi]=\int{\rm d}^3x\,\left\{\frac{c}{2}(\nabla\phi)^2+
\frac{\kappa}{2}(\nabla^2\phi)^2+g(\phi({\bf x}))\right\}\,,
\label{eeq1}
\ee
where $c,\kappa>0$ are ``stiffness'' parameters and $g(\phi)$ is the specific
free energy (i.e., Gibbs free energy per unit volume) of the homogeneous
system, the bulk liquid being the reference state where $\phi=0$.
Exactly at coexistence, the OP values are
$\phi_-=\phi_{s0}$ in the bulk solid and $\phi_+=0$ in the bulk liquid
(i.e., $g(\phi_{s0})=g(0)=0$ and $g(\phi)>0$ otherwise).
When boundary conditions are applied such that $\phi\rightarrow\phi_\pm$
for $z\rightarrow\pm\infty$, a planar interface orthogonal to $z$
is forced to appear in the system. The corresponding $z$-dependent OP profile
is the extremal point $\phi_0(z)$ of (\ref{eeq1}) that satisfies
the boundary conditions:
\be
c\phi_0^{\prime\prime}-\kappa\phi_0^{\prime\prime\prime\prime}=
\frac{{\rm d}g}{{\rm d}\phi}(\phi_0;T=T_m)\,,\,\,\,\,\,\,{\rm with}\,\,\,
\phi_0(-\infty)=\phi_{s0}\,\,\,{\rm and}\,\,\,\phi_0(+\infty)=0\,.
\label{eeq2}
\ee
Evidently, ${\cal G}[\phi_0]$ represents the free-energy cost of the
solid-liquid interface.

Away from coexistence, the absolute minimum of $g(\phi)$ falls at
$\phi=\phi_s>0$ for $\Delta T\equiv T-T_m<0$. This can be described by
\be
g(\phi)=c_2\phi^2+c_3\phi^3+c_4\phi^4+\ldots
\label{eeq3}
\ee
if we take $c_2=c_{20}+c'_{20}\Delta T$ ($c_{20},c'_{20}>0$), all other
$c_n$ coefficients being constant. Not far from $T_m$, the OP profile
of a spherical solid cluster of radius $R\gg\sqrt{2c/c_{20}}$ is well
described by $\phi_0(r-R)$, provided that the center of $\phi_0(z)$ is
chosen at $z=0$. Based on these assumptions, the free energy of cluster
formation becomes~\cite{SMnoi}:
\be
\Delta G(R)=4\pi\int_0^{+\infty}{\rm d}r\,r^2\left[c\phi_0^{\prime\,2}(r-R)+
2\kappa\phi_0^{\prime\prime\,2}(r-R)\right]-4\pi c'_{20}|\Delta T|
\int_0^{+\infty}{\rm d}r\,r^2\phi_0^2(r-R)\,.
\label{eeq4}
\ee
A straightforward calculation then shows that:
\ba
&& \int_0^{+\infty}{\rm d}r\,r^2\phi_0^2(r-R)=\frac{1}{3}\phi_{s0}^2R^3+
(I'_2-I_2)+2(I'_1+I_1)R+(I'_0-I_0)R^2\,;
\nonumber \\
&& \int_0^{+\infty}{\rm d}r\,r^2\phi_0^{\prime\,2}(r-R)=J_2+2J_1R+J_0R^2\,;
\nonumber \\
&& \int_0^{+\infty}{\rm d}r\,r^2\phi_0^{\prime\prime\,2}(r-R)=
K_2+2K_1R+K_0R^2\,,
\label{eeq5}
\ea
where, for $n=0,1,2$:
\ba
I_n&=&\int_0^{+\infty}{\rm d}z\,z^n\left(\phi_{s0}^2-\phi_0^2(-z)\right)\,\,\,
{\rm and}\,\,\,I'_n=\int_0^{+\infty}{\rm d}z\,z^n\phi_0^2(z)\,;
\nonumber \\
J_n&=&\int_{-\infty}^{+\infty}{\rm d}z\,z^n\phi_0^{\prime\,2}(z)\,;
\nonumber \\
K_n&=&\int_{-\infty}^{+\infty}{\rm d}z\,z^n\phi_0^{\prime\prime\,2}(z)\,.
\label{eeq6}
\ea
Substituting Eqs.\,(\ref{eeq5}) into (\ref{eeq4}), we obtain the final
expression for $\Delta G(R)$ (Eq.\,(2) of MT, with $C=0$):
\be
\Delta G(R)=4\pi R^2\widetilde{\sigma}\left(1-\frac{2\widetilde{\delta}}{R}+
\frac{\widetilde{\epsilon}}{R^2}\right)-\frac{4}{3}\pi R^3\rho_s|\Delta\mu|\,,
\label{eeq7}
\ee
where
\ba
&& \widetilde{\sigma}=cJ_0+2\kappa K_0-c'_{20}(I'_0-I_0)|\Delta T|\,;
\nonumber \\
&& \widetilde{\sigma}\widetilde{\delta}=-cJ_1-2\kappa K_1+
c'_{20}(I'_1+I_1)|\Delta T|\,;
\nonumber \\
&& \widetilde{\sigma}\widetilde{\epsilon}=cJ_2+2\kappa K_2-
c'_{20}(I'_2-I_2)|\Delta T|\,;
\nonumber \\
&& \rho_s\Delta\mu=c'_{20}\phi_{s0}^2\Delta T\,.
\label{eeq8}
\ea
In particular, the solid-liquid interface tension $\sigma$ and the
``Tolman length'' $\delta$ are given by:
\ba
&& \sigma\equiv cJ_0+2\kappa K_0=\int_{-\infty}^{+\infty}{\rm d}z
\left[c\phi_0^{\prime\,2}(z)+2\kappa\phi_0^{\prime\prime\,2}(z)\right]\,;
\nonumber \\
&& \delta\equiv-\frac{1}{\sigma}\left(cJ_1+2\kappa K_1\right)=
-\frac{\int_{-\infty}^{+\infty}{\rm d}z\,z
\left[c\phi_0^{\prime\,2}(z)+2\kappa\phi_0^{\prime\prime\,2}(z)\right]}
{\int_{-\infty}^{+\infty}{\rm d}z
\left[c\phi_0^{\prime\,2}(z)+2\kappa\phi_0^{\prime\prime\,2}(z)\right]}\,.
\label{eeq9}
\ea
Note that, for $\kappa=0$, the formula for $\delta$ reduces to
that reported in \cite{SMFisher}.
A nonzero $\delta$ corresponds to a $\phi_0(z)$ that is not symmetric
around 0, namely to an interface between phases of a different nature. 
Summing up, Eq.\,(\ref{eeq7}) describes the corrections to the classical
nucleation theory (CNT) which arise by replacing the assumption of a
sharp solid-liquid interface with a more realistic finite width.

%
%
\section{Calculation of the surface tension and the Tolman length}

Given the form of $g$, one can compute the explicit values of $\sigma$ and
$\delta$ in Eqs.\,(\ref{eeq9}), and of $\epsilon=(cJ_2+2\kappa K_2)/\sigma$
(i.e., the value of $\widetilde{\epsilon}$ at $T_m$, see Eq.\,(\ref{eeq8})),
once the exact $\phi_0(z)$ is known. In turn, $\phi_0(z)$ follows from solving
the boundary value problem (\ref{eeq2}), which can be simplified, after an
integration by parts, to
\be
\kappa\phi_0'\phi_0'''=\frac{c}{2}\phi_0'^2+\frac{\kappa}{2}\phi_0''^2-
g(\phi_0)\,,\,\,\,\,\,\,{\rm with}\,\,\,
\phi_0(-\infty)=\phi_{s0}\,\,\,{\rm and}\,\,\,\phi_0(+\infty)=0\,.
\label{eeq10}
\ee
A special case of $g$ function is used in the MT, where at $T_m$ we
take
\be
g(\phi)=c_{20}\phi^2\left(1-\frac{\phi}{\phi_{s0}}\right)^2\left(1+\tau
\frac{\phi}{\phi_{s0}}\right)\equiv
g_0(\phi)\left(1+\tau\frac{\phi}{\phi_{s0}}\right)\,.
\label{eeq11}
\ee
For this $g$, the differential equation (\ref{eeq10}) is still too difficult
to solve in closed form for generic $\kappa$, even for $\tau=0$. Hence, we
decide to work perturbatively in $\kappa$ and $\tau$.

At zeroth order, i.e., $\kappa=\tau=0$, corresponding to $\phi^4$ theory,
the solution to (\ref{eeq10}) is
\be
\bar{\phi}_0(z)=\frac{\phi_{s0}}{2}\left\{1-\tanh\left(\frac{z-C}{\ell}\right)
\right\}
\label{eeq12}
\ee
with $\ell=\sqrt{2c/c_{20}}$ and arbitrary $C$. We fix $C$ by requiring that
the interface is centered at $z=0$, i.e., by imposing
\be
\int_{-\infty}^{+\infty}{\rm d}z\,z\bar{\phi}_0'(z)=0
\label{eeq13}
\ee
(hence $C=0$). Next, we take non-zero $\kappa$ and $\tau$, assumed to be
of the same order of magnitude, and search for a first-order solution to
(\ref{eeq10}) in the form
\be
\phi_0(z)=\bar{\phi}_0(z)+\tau\psi_1(z)+\frac{\kappa}{c\ell^2}\chi_1(z)\,.
\label{eeq14}
\ee
Upon inserting this function into Eq.\,(\ref{eeq10}), we obtain two independent
equations for $\psi_1(z)$ and $\chi_1(z)$, namely
\be
c\bar{\phi}_0'\psi_1'-g_0'(\bar{\phi}_0)\psi_1=
\frac{\bar{\phi}_0g_0(\bar{\phi}_0)}{\phi_{s0}}
\label{eeq15}
\ee
and
\be
c\bar{\phi}_0'\chi_1'-g_0'(\bar{\phi}_0)\chi_1=c\ell^2\left(
\bar{\phi}_0'\bar{\phi}_0'''-\frac{1}{2}\bar{\phi}_0''^2\right)\,.
\label{eeq16}
\ee
The solutions to Eqs.\,(\ref{eeq15}) and (\ref{eeq16}) such that each term
in Eq.\,(\ref{eeq14}) separately meets a requirement analog to (\ref{eeq13})
are:
\be
\psi_1(z)=-\frac{\phi_{s0}}{8\cosh^2(z/\ell)}\left(1-\ln 2+
\frac{z}{\ell}-\ln\cosh\frac{z}{\ell}\right)
\label{eeq17}
\ee
and
\be
\chi_1(z)=\frac{\phi_{s0}}{\cosh^2(z/\ell)}
\left(2\tanh\frac{z}{\ell}-\frac{z}{\ell}\right)\,.
\label{eeq18}
\ee
Upon plugging the by now specified $\phi_0(z)$ in the integrals defining
$\sigma,\delta$, and $\epsilon$, we eventually obtain the final expressions
quoted in Eq.\,(4) of MT.

%
%
\section{Derivation of the interface Hamiltonian}

We synthetically show how the interface Hamiltonian, Eq.\,(5) of MT,
can be obtained from the Landau free energy (\ref{eeq1}).
For this derivation, we build on Refs.\,\cite{SMKogon,SMKassner}.
More details will be given in \cite{SMnoi}.

In the same spirit of a statistical field theory for vesicles,
we wish to assign a free energy cost to each particular realization
of the cluster interface, here assumed to be sharp and akin
to a mathematical surface $\Sigma$. The goal here is to describe
the effect of shape fluctuations --- though, in practice,
in order to make analytical progress, we shall be forced in the next
paragraph to describe just small deviations around the spherical shape.

Let ${\bf R}(u,v)$ be the parametrization (``coordinate patch'') of a tiny
piece of $\Sigma$. For points
${\bf r}$ close to this small portion of $\Sigma$, we switch from 3D cartesian
coordinates, ${\bf r}=(x,y,z)$, to new coordinates $q_\alpha=(u,v,\zeta)$
(tangential and normal to $\Sigma$):
\be
{\bf r}={\bf R}(u,v)+\zeta\widehat{\bf n}(u,v)\,,
\label{eeq19}
\ee
where
\be
\widehat{\bf n}(u,v)=
\frac{{\bf R}_u\wedge{\bf R}_v}{|{\bf R}_u\wedge{\bf R}_v|}
\label{eeq20}
\ee
is the unit normal to $\Sigma$. For a patch that deviates only slightly from
planarity, we may adopt a free energy of ${\cal G}[\phi_0(\zeta(x,y,z))]$,
thus arriving, by Eq.(\ref{eeq1}), to the surface Hamiltonian
\be
{\cal H}_s=\int{\rm d}u\,{\rm d}v\,{\rm d}\zeta\,|J|\left\{\frac{c}{2}
\left(\nabla\phi_0(\zeta)\right)^2+\frac{\kappa}{2}
\left(\nabla^2\phi_0(\zeta)\right)^2+g(\phi_0(\zeta))\right\}
\label{eeq21}
\ee
with $J=|{\bf r}_u\cdot({\bf r}_v\wedge{\bf r}_\zeta)|=
\widehat{\bf n}\cdot({\bf r}_u\wedge{\bf r}_v)$. In order to compute
the explicit form of $J$ as well as those of the gradient and Laplacian of
a function of $\zeta$ only, it is convenient to take a patch parametrization
in terms of orthonormal, arc-length coordinates, i.e., one such that
${\bf R}_u\cdot{\bf R}_v=0$ all over the patch and
$|{\bf R}_u|=|{\bf R}_v|=1$. Although this can rigorously be done only for
surfaces with zero Gaussian curvature ($K=0$)~\cite{SMAbate}, we can reasonably
expect to make small errors of order $K$ for quasi-planar interfaces.
It then follows that
\ba
\frac{\partial{\bf r}}{\partial\zeta} &=& \widehat{\bf n}\,;
\nonumber \\
\frac{\partial{\bf r}}{\partial u} &=& (1-\zeta\kappa_n^{(1)}){\bf R}_u-
\zeta\tau_g{\bf R}_v\,;
\nonumber \\
\frac{\partial{\bf r}}{\partial v} &=& -\zeta\tau_g{\bf R}_u+
(1-\zeta\kappa_n^{(2)}){\bf R}_v\,,
\label{eeq22}
\ea
where $\kappa_n^{(1)}$ and $\kappa_n^{(2)}$ are the normal curvatures of
the $u$- and $v$-lines respectively, and $\tau_g\equiv\tau_g^{(1)}=
-\tau_g^{(2)}$ is the geodetic torsion.
From Eqs.\,(\ref{eeq22}), one can derive the metric tensor
$g_{\alpha\beta}$ of the transformation (\ref{eeq19}) and the Jacobian:
\be
g_{\alpha\beta}\equiv\frac{\partial{\bf r}}{\partial q_\alpha}\cdot
\frac{\partial{\bf r}}{\partial q_\beta}
=\left(\begin{array}{ccc}
\left(1-\zeta\kappa_n^{(1)}\right)^2+\zeta^2\tau_g^2 & -2\zeta\tau_g+
\zeta^2\tau_g\left(\kappa_n^{(1)}+\kappa_n^{(2)}\right) & 0 \\
-2\zeta\tau_g+\zeta^2\tau_g\left(\kappa_n^{(1)}+\kappa_n^{(2)}\right)
& \left(1-\zeta\kappa_n^{(2)}\right)^2+\zeta^2\tau_g^2 & 0 \\
0 & 0 & 1
\end{array}\right)
\label{eeq23}
\ee
and
\be
J=\left(1-\zeta\kappa_n^{(1)}\right)\left(1-\zeta\kappa_n^{(2)}\right)
-\zeta^2\tau_g^2\equiv\sqrt{g}\,,
\label{eeq24}
\ee
$g$ being the determinant of (\ref{eeq23}).
Considering that covariant and contravariant components of a vector
${\bf v}=\sum_{i=1}^3\left({\bf v}\cdot\widehat{\bf x}_i\right)
\widehat{\bf x}_i$ are built by projecting {\bf v} on the bases
$\nabla q_\alpha$ and $\partial{\bf r}/\partial q_\alpha$, respectively,
we can calculate the gradient of a scalar field $\phi$ and the divergence
of a vector field {\bf A} in local coordinates as follows:
\be
\nabla\phi=\frac{\partial\phi}{\partial q_\alpha}g^{\alpha\beta}
\frac{\partial{\bf r}}{\partial q_\beta}\,\,\,\,\,\,{\rm and}\,\,\,\,\,\,
\nabla\cdot{\bf A}=\frac{1}{\sqrt{g}}\frac{\partial}{\partial q_\alpha}
\left(\sqrt{g}A^\alpha\right)\,,
\label{eeq25}
\ee
$g^{\alpha\beta}$ being the inverse of (\ref{eeq23}). In particular,
\be
\nabla\phi(\zeta)=\phi^{\prime}(\zeta)\widehat{\bf n}\,\,\,\,\,\,{\rm and}
\,\,\,\,\,\,\nabla^2\phi(\zeta)=\phi^{\prime\prime}(\zeta)+\phi^{\prime}(\zeta)
\nabla\cdot\widehat{\bf n}\,,
\label{eeq26}
\ee
where
\be
\nabla\cdot\widehat{\bf n}=\frac{1}{\sqrt{g}}
\left(-\kappa_n^{(1)}-\kappa_n^{(2)}-2\zeta\tau_g^2\right)\,.
\label{eeq27}
\ee
Finally, the mean and Gaussian curvatures of the patch are given by
\be
H=\left.\frac{1}{2}\nabla\cdot\widehat{\bf n}\right|_{\zeta=0}=
-\frac{1}{2}\left(\kappa_n^{(1)}+\kappa_n^{(2)}\right)
\label{eeq28}
\ee
and
\be
K=\left.\widehat{\bf n}\cdot\left(\frac{\partial\widehat{\bf n}}
{\partial u}\wedge\frac{\partial\widehat{\bf n}}{\partial v}\right)
\right|_{\zeta=0}=\kappa_n^{(1)}\kappa_n^{(2)}-\tau_g^2\,.
\label{eeq29}
\ee

We are now in a position to simplify Eq.\,(\ref{eeq21}).
Upon using Eq.\,(\ref{eeq2}) to eliminate $g(\phi_0)$ in favor of
$(c/2)\phi_0^{\prime\,2}-\kappa\left[
\left(\phi_0^{\prime}\phi_0^{\prime\prime}\right)^{\prime}-
(3/2)\phi_0^{\prime\prime\,2}\right]$, and inserting
Eqs.\,(\ref{eeq24}), (\ref{eeq26}), (\ref{eeq28}), and (\ref{eeq29}),
we eventually get~\cite{SMnoi} the classic Canham-Helfrich Hamiltonian
for fluid membranes,
\be
{\cal H}_s[\Sigma]=\int_\Sigma{\rm d}S\,\left(\bar{a}+\bar{b}H+\bar{c}H^2+
\bar{d}K\right)\,,
\label{eeq30}
\ee
with the following explicit expressions for the coefficients:
\ba
\bar{a} &=& \int_{-\infty}^{+\infty}{\rm d}\zeta
\left[c\phi_0^{\prime\,2}(\zeta)+2\kappa
\phi_0^{\prime\prime\,2}(\zeta)\right]\,;
\nonumber \\
\bar{b} &=& 2\int_{-\infty}^{+\infty}{\rm d}\zeta\,\zeta
\left[c\phi_0^{\prime\,2}(\zeta)+2\kappa
\phi_0^{\prime\prime\,2}(\zeta)\right]\,;
\nonumber \\
\bar{c} &=& 2\kappa\int_{-\infty}^{+\infty}{\rm d}\zeta\,
\phi_0^{\prime\,2}(\zeta)\,;
\nonumber \\
\bar{d} &=& \int_{-\infty}^{+\infty}{\rm d}\zeta\left\{\zeta^2
\left[c\phi_0^{\prime\,2}(\zeta)+2\kappa
\phi_0^{\prime\prime\,2}(\zeta)\right]-\kappa
\phi_0^{\prime\,2}(\zeta)\right\}\,.
\label{eeq31}
\ea
A few remarks are now in order:
a) in deriving Eq.\,(\ref{eeq30}), all subleading corrections to the
$H^2$ and $K$ terms were ignored.
b) $H$ and $K$ are reparametrization invariants, hence no ambiguity
arises from the arbitrariness of the parametrization used.
c) The above derivation actually applies for just one $\Sigma$ patch.
However, upon viewing $\Sigma$ as the union of many disjoint patches, 
the Hamiltonian (\ref{eeq30}) holds for $\Sigma$ as well.
d) As anticipated, the coefficient $\bar{d}$ of the $K$ term in (\ref{eeq30})
is possibly different from the quoted one since a parametrization in terms
of orthonormal coordinates does not generally exist. However, as far as
we only allow for clusters with the topology of a sphere,
$\int_\Sigma{\rm d}S\,K$ takes the constant value of $4\pi$
by the Gauss-Bonnet theorem and the $K$ term in ${\cal H}_s$ can be dropped.
Comparing the definition of $\bar{a}$ and $\bar{b}$ in Eqs.\,(\ref{eeq31})
with Eq.\,(\ref{eeq9}), we can rewrite Eq.\,(\ref{eeq30}) in the form
\be
{\cal H}_s[\Sigma]=\int_\Sigma{\rm d}S\,\left(\sigma-2\sigma\delta H+
2\lambda H^2\right)\,,
\label{eeq32}
\ee
where $\lambda=\bar{c}/2$.
e) The term linear in $H$ is related to the {\em spontaneous curvature} of
$\Sigma$, $H_0=-\bar{b}/(2\bar{c})$, which is proportional to the Tolman length
$\delta$. A nonzero value of $H_0$ entails a difference in energy between
inward and outward interface protrusions, with the effect of producing
a nonzero Tolman length $\delta$. The realization that in systems where,
as in the Ising model, the symmetry is perfect between the two phases then
$\delta=0$, has long been known~\cite{SMFisher}.

%
%
\section{Field theory for the interface Hamiltonian}

We consider a single solid cluster in a supercooled-liquid host.
We model the cluster surface $\Sigma$ as being subject to random
fluctuations distributed according to
$\exp\{-\beta{\cal H}_s[\Sigma]\}$ with $\beta=1/(k_BT)$
and ${\cal H}_s[\Sigma]$ given as in Eq.\,(5) of MT.
We stress that this form of interface Hamiltonian is only valid slightly
below $T_m$, where the curvature of the cluster surface is small.
The cost in free energy of a cluster of volume $V$ is then taken to be
\be
\Delta G(V)=-\rho_s|\Delta\mu|V+F_s(V)\,,
\label{eeq33}
\ee
where the surface free energy $F_s=-(1/\beta)\ln Z_s$ with
\be
Z_s(V)=a^3\int{\cal D}\Sigma\,\delta({\cal V}[\Sigma]-V)
\exp\{-\beta{\cal H}_s[\Sigma]\}\,.
\label{eeq34}
\ee
In the above equation, $a=\rho_s^{-1/3}$ is a characteristic system length,
${\cal V}[\Sigma]$ is the volume enclosed by $\Sigma$, and
${\cal D}\Sigma$ is a suitable integral measure (see below).
Neglecting overhangs and liquid inclusions in the cluster, let
$r=R(\theta,\phi)$ be the equation of $\Sigma$ in spherical coordinates.
To proceed further, we assume only small deviations from a sphere, i.e.,
$R(\theta,\phi)=R_0[1+\epsilon(\theta,\phi)]$, with
$\epsilon(\theta,\phi)\ll 1$.
Then, we expand $\epsilon(\theta,\phi)$ in real spherical harmonics,
\be
\epsilon(\theta,\phi)=
\sum_{l=1}^\infty\sum_{m=-l}^lx_{l,m}Y_{l,m}(\theta,\phi)\,,
\label{eeq35}
\ee
and we agree to ignore, from now on, all terms beyond second-order in the
coefficients $x_{l,m}$. With these specifications, the enclosed volume and
area of $\Sigma$ are approximated as
\be
{\cal V}[\Sigma]=\frac{4}{3}\pi R_0^3+R_0^3\sum_{l>0,m}x_{l,m}^2\equiv
\frac{4}{3}\pi R_0^3\,f(\{x\})
\label{eeq36}
\ee
and
\be
{\cal A}[\Sigma]=4\pi R_0^2
+\frac{R_0^2}{2}\sum_{l>0,m}\left(l^2+l+2\right)x_{l,m}^2\equiv
4\pi R_0^2\,g(\{x\})\,,
\label{eeq37}
\ee
respectively, $f(\{x\})$ and $g(\{x\})$ being close-to-1 factors.
In order to evaluate the mean curvature $H$, we move from
\be
\nabla\cdot\widehat{\bf n}=\frac{2}{R(\theta,\phi)}
\left(1+\frac{1}{2}L^2\epsilon(\theta,\phi)-
\frac{1}{2}\epsilon(\theta,\phi)L^2\epsilon(\theta,\phi)\right)\,,
\label{eeq38}
\ee
where
\be
L^2=-\frac{1}{\sin\theta}\frac{\partial}{\partial\theta}\left(\sin\theta
\frac{\partial}{\partial\theta}\right)-
\frac{1}{\sin^2\theta}\frac{\partial^2}{\partial\phi^2}\,.
\label{eeq39}
\ee
Eventually, we obtain~\cite{SMnoi}:
\ba
&& \int_\Sigma{\rm d}S\left(\sigma-2\sigma\delta H+2\lambda H^2\right)=
4\pi\sigma R_0^2+\frac{\sigma R_0^2}{2}\sum_{l>0,m}(l^2+l+2)x_{l,m}^2
\nonumber \\
&& -8\pi\sigma\delta R_0-\sigma\delta R_0\sum_{l>0,m}l(l+1)x_{l,m}^2+
8\pi\lambda+\frac{\lambda}{2}\sum_{l>1,m}l(l+1)(l-1)(l+2)x_{l,m}^2\,.
\label{eeq40}
\ea
Finally, we specify the integral measure in (\ref{eeq34}):
\be
\int{\cal D}\Sigma=\int_{-\infty}^{+\infty}\prod_{l>0,m}\left(
\frac{S}{s}\,{\rm d}x_{l,m}\right)\int_0^{+\infty}
\frac{{\rm d}R_0}{a}\,,
\label{eeq41}
\ee
where $S=(36\pi)^{1/3}V^{2/3}$ is the area of the spherical surface of
volume $V$ and $s=4\pi a^2$. Equation (\ref{eeq41}) follows from
requiring that the present theory be the continuum limit of the field
theory for a solid-on-solid (SOS) model whose heights are defined on
nodes uniformly placed over a sphere of radius
$\sqrt{S/(4\pi)}$~\cite{SMnoi}.

We can now compute the partition function (\ref{eeq34}). We calculate first
the integral on $R_0$ by rearranging the delta function in $Z_s$ as
\be
\delta\left(\frac{4}{3}\pi R_0^3\,f(\{x\})-V\right)=
\frac{\delta\left(R_0-\left[4\pi f(\{x\})/(3V)\right]^{-1/3}\right)}
{(36\pi)^{1/3}V^{2/3}f(\{x\})^{1/3}}\,.
\label{eeq42}
\ee
After doing the trivial integral over $R_0$, we remain with a factor
$f(\{x\})^{-1/3}$ which, within a quadratic theory, can be treated as follows:
\be
f(\{x\})^{-1/3}=\left(1+\frac{3}{4\pi}\sum_{l>0,m}x_{l,m}^2\right)^{-1/3}
\simeq 1-\frac{1}{4\pi}\sum_{l>0,m}x_{l,m}^2\simeq
\exp\left\{-\frac{1}{4\pi}\sum_{l>0,m}x_{l,m}^2\right\}\,.
\label{eeq43}
\ee
In the end, we arrive at a Gaussian integral which is readily computed:
\ba
Z_s &=& (36\pi)^{-1/3}\left(\frac{V}{a^3}\right)^{-2/3}
\exp\left\{\beta\rho_s|\Delta\mu|V-\beta\sigma S-8\pi\beta\lambda+
8\pi\beta\sigma\delta\left(\frac{3V}{4\pi}\right)^{1/3}\right\}
\nonumber \\
&\times& \left(\frac{2\pi S}{s}\right)^3
\prod_{l>1}\left\{\left(\frac{s}{2\pi S}\right)^2
\left[1+\frac{\beta\sigma S}{2}(l^2+l-2)+
2\pi\beta\lambda\,l(l+1)(l-1)(l+2)\right.\right.
\nonumber \\
&-& \left.\left.4\pi\beta\sigma\delta
\sqrt{\frac{S}{4\pi}}(l^2+l-2)\right]\right\}^{-(l+1/2)}\,.
\label{eeq44}
\ea
Without a proper ultraviolet cutoff $l_{\rm max}$, the $l$ sum in
$F_s=-k_BT\ln Z_s$ does not converge. This is a typical occurrence for
field theories on the continuum, which do not take into account the
granularity of matter at the most fundamental level.
We fix $l_{\rm max}$ by requiring that the total number of $(l,m)$
modes be equal to the average number of SOS heights/atoms on the cluster
surface. It thus follows:
\be
l_{\rm max}=\frac{\sqrt{S}}{a}-1\,.
\label{eeq45}
\ee
With this cutoff, the surface free energy becomes $F_s=\gamma(S)S$,
with a surface tension $\gamma(S)$ dressed by thermal fluctuations:
\ba
\gamma(S) &=& \sigma+\frac{k_BT}{2S}\sum_{l=2}^{\sqrt{S}/a-1}(2l+1)
\ln\left[A+B(l^2+l-2)+C(l^2+l-2)^2\right]
\nonumber \\
&-& 2\sigma\delta\left(\frac{4\pi}{S}\right)^{1/2}
-2\frac{k_BT}{S}\ln\left(\frac{S}{a^2}\right)
-3\frac{k_BT}{S}\ln\left(\frac{2\pi a^2}{s}\right)+
\frac{8\pi\lambda}{S}\,.
\label{eeq46}
\ea
The quantities $A,B$, and $C$ in Eq.\,(\ref{eeq46}) are given by
\be
A=\frac{A_0}{S^2}\,,\,\,\,B=\frac{2C_0}{S^2}+\frac{D_0}{S\sqrt{S}}+
\frac{B_0}{S}\,,\,\,\,C=\frac{C_0}{S^2}\,,
\label{eeq47}
\ee
where
\be
A_0=\frac{s^2}{4\pi^2}\,,\,\,\,B_0=\frac{\beta\sigma s^2}{8\pi^2}\,,
\,\,\,C_0=\frac{\beta\lambda s^2}{2\pi}\,,\,\,\,
D_0=-\frac{\beta\sigma\delta s^2}{2\pi\sqrt{\pi}}\,.
\label{eeq48}
\ee
By the Euler-Mac Laurin formula, the residual sum in Eq.\,(\ref{eeq46})
can be evaluated explicitly. After a tedious and rather lengthy derivation,
we obtain (for $\lambda\neq 0$):
\ba
&& \gamma(S)=\sigma+\frac{k_BT}{2a^2}\left[\ln\frac{B_0}{a^2e^2}+
\left(1+\frac{B_0a^2}{C_0}\right)\ln\left(1+\frac{C_0}{B_0a^2}\right)\right]
\nonumber \\
&& +\left[-2\sigma\delta+\frac{k_BTD_0}{4C_0\sqrt{\pi}}
\ln\left(1+\frac{C_0}{B_0a^2}\right)\right]\left(\frac{4\pi}{S}\right)^{1/2}-
\frac{7}{6}k_BT\frac{\ln(S/a^2)}{S}
\nonumber \\
&& +\left[8\pi\beta\lambda-3\ln\frac{2\pi a^2}{s}-
\frac{11}{6}\ln\frac{B_0}{a^2}+3-\frac{5}{3}\ln 2-\frac{25}{96}+
\frac{121}{46080}\right.
\nonumber \\
&& +\frac{D_0a}{4C_0}-\frac{D_0^2}{4B_0C_0}-
\frac{1}{6}\ln\left(\frac{B_0}{a^2}+\frac{C_0}{a^4}\right)
+\frac{1}{8C_0(B_0a^2+C_0)^2}
\nonumber \\
&& \times\left(
-4B_0C_0D_0a^3-18B_0C_0^2a^2-2C_0^2D_0a-\frac{28}{3}C_0^3-
\frac{26}{3}B_0^2C_0a^4\right.
\nonumber \\
&& \left.\left.-2B_0^2D_0a^5+2B_0D_0^2a^4+2C_0D_0^2a^2
\right)\right]\frac{k_BT}{S}\,,
\label{eeq49}
\ea
up to terms $o(S^{-1})$.
We wrote a computer code to evaluate the sum in (\ref{eeq46}) numerically,
and so checked that every single term in Eq.\,(\ref{eeq49}) is correct.
In the notation of MT, the interface free energy $\gamma(S)$ has the form:
\be
\gamma=\widetilde{\sigma}\left(1-\frac{2\widetilde{\delta}}{R}+
\frac{\widetilde{\epsilon}}{R^2}\right)-
\frac{7}{3}k_BT\frac{\ln(R/a)}{4\pi R^2}\,,
\label{eeq50}
\ee
with $R=[3V/(4\pi)]^{1/3}$ and $T$-dependent expressions for
$\widetilde{\sigma},\widetilde{\delta}$, and $\widetilde{\epsilon}$
that can be read in Eq.\,(\ref{eeq49}).
At coexistence, the values of $\widetilde{\sigma}$ and $\widetilde{\delta}$
for a quasi-spherical cluster differ from those in Landau theory (i.e.,
$\sigma$ and $\delta$) for fluctuation corrections which are the effect
of thermally excited capillary waves on the cluster surface.


\begin{thebibliography}{99}
\bibitem{nota} If crystallinity values are restricted between 0 and
$\phi_{s0}$, one can pass over the fact that this $g(\phi)$ function
is not bounded from below.
\bibitem{footnote} Our fresh data for $h=0.55$ seem to differ from those in
Fig.\,5 of Ref.\,\cite{Pan} by just a constant --- apparently, the authors of
\cite{Pan} have set the zero of $\Delta G(n)$ at $n=1$.
\end{thebibliography}
\end{document}